\documentclass[runningheads,fleqn]{svmult}
\usepackage{makeidx}   
\usepackage{graphicx}  
\usepackage{subeqnar}  
\usepackage{multicol}  
\usepackage{taphys}    
\makeindex             
\begin{document}
\title*{Applications of adiabatic passage \\ in solid-state devices}
\toctitle{Applications of adiabatic passage \protect\newline 
          in solid-state devices}
%
%
\titlerunning{Applications of adiabatic passage in solid-state devices}
%
\author{Jens Siewert\inst{1}
\and Tobias Brandes\inst{2}}
\authorrunning{J.\ Siewert and T.\ Brandes}
%
%
\institute{Institut f\"ur Theoretische Physik, Universit\"at Regensburg,
           D-93040 Regensburg, Germany
\and Department of Physics, The University of Manchester, 
     Manchester M60 1QD, United Kingdom}

\maketitle              

\begin{abstract}
Coherent population transfer by adiabatic passage is a
well-known method in quantum optics. This remarkable technique which is
based on simple ideas has remained largely unknown to solid-state
physicists. Here we provide an introduction to the basic principles
of this method and discuss also some applications in solid-state
systems.\\[1mm]
PACS number(s): 32.80.Qk, 73.23.-b, 73.40.Gk
\end{abstract}

\section{Introduction}
Traditionally solid-state physics is focused on systems with 
quasi-continuous energy spectra for the various elementary 
excitations present in macroscopic solids. During the past
decade, however, the successful development of nano\-technology
has made it possible to study devices below the micrometer
scale, to manipulate single charges and fluxes, or to build
small electrical and mechanical resonators. The typical energy 
scales for such
devices are, say, on the order of \mbox{0.1 \ldots 1 Kelvin} and 
therefore much 
larger than the smallest temperatures that can be reached in
an experiment. Consequently, the discrete nature of the energy
spectrum needs to be taken into account in order to understand
the behavior of these devices.

Another interesting aspect of solid-state systems at the nanoscale
is that they may exhibit quantum coherence phenomena. These new
effects have attracted significant attention in the recent past.
Controlling quantum coherence in solid-state nanodevices is one
of the major objectives in present-day research. Achievement of
this goal would make it possible to study quantum dynamics on
a macroscopic scale, to provide better insight into the mechanisms
of decoherence, and possibly to realize 
quantum information processing in practice.

A consequence of this evolution is that methods used, e.g., in quantum
optics become directly relevant also in solid-state physics.
A particularly interesting technique is the so-called stimulated 
Raman adiabatic passage (STIRAP) that has been developed by
Bergmann and co-workers \cite{BergBuch,BergRMP}. 
This method can be used to change the quantum state of a system 
by controlling certain coupling parameters.  The mathematics
underlying this technique is rather simple and general such that
it may apply to physically completely different situations. 

Until now, adiabatic passage has had only very few applications in
solid-state devices, e.g., \cite{Tobias1,Tobias2,NonAbel,Alec}.
However, one may hope that this method, due to its elegance 
and simplicity, will attract more interest and find new 
applications to solid-state systems in the near future.
In this contribution we provide a brief introduction to the
basic principles of adiabatic passage. In Section 2 we will
explain population transfer in three-level atoms, as it is
known from quantum optics. In Section 3,
the method is `translated' to a simple solid-state device, 
a superconducting Cooper-pair box with three islands. We will show
that adiabatic charge transfer between the islands is possible
in close analogy to the three-level atom. Finally, in Section 4
we will briefly discuss two advanced applications that are based
on the technique of adiabatic passage. In our opinion, these 
applications may serve to illustrate the enormous potential of 
the method in the field of condensed-matter physics.

\section{Adiabatic Passage in Three-Level Atoms}
In this section, we briefly discuss the basics of adiabatic passage.
A more complete discussion of the underlying physics
can be found, e.g., 
in the review articles Refs.\ \cite{BergBuch,BergRMP} and 
in the textbook Ref.\ \cite{Scully}.

\begin{figure}
\begin{center}
\includegraphics[width=.6\textwidth]{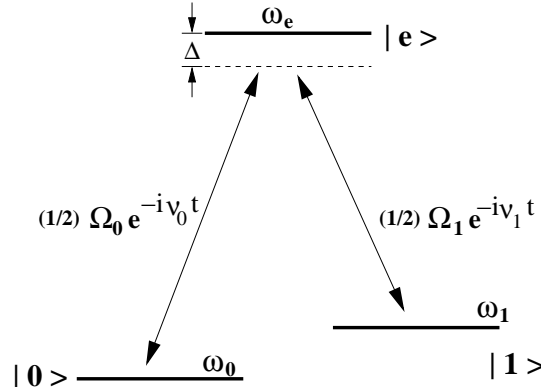}
\end{center}
\vspace*{-3mm}
\caption[]{Three-level atom with two ground states $|0\rangle$, $|1\rangle$
           coupled to an excited state $|e\rangle$ via laser fields
           $(1/2)\Omega_0 \exp{(-i\nu_0 t)}$, $(1/2)\Omega_1 \exp{(-i\nu_1 t)}$.
           Note that there is no direct coupling between the states
           $|0\rangle$ and $|1\rangle$.
           }
\label{fig1}
\end{figure}

Consider an atom with a $\Lambda$-type three-level configuration
as shown in \mbox{Fig.\ \ref{fig1}}. The long-lived ground states $| 0 \rangle$
and $| 1 \rangle$ (energies $\omega_0$, $\omega_1$) are coupled to 
an excited state $| e \rangle$ (energy $\omega_e$) via (classical) laser fields 
with Rabi frequencies 
$\Omega_0$, $\Omega_1$. 
The laser frequencies are assumed to have the same detuning $\Delta$
with respect to the atomic transitions
$$
\Delta=(\omega_e-\omega_0)-\nu_0=(\omega_e-\omega_1)-\nu_1\ \ .
$$
\vspace*{-6mm}

If the state $|\psi(t)\rangle$ of the atom
is written in the form
\begin{equation}
     |\psi(t)\rangle\ =\ 
                         c_e(t)e^{-i(\omega_e-\Delta) t}|e\rangle\ +\ 
                         c_0(t)e^{-i\omega_0 t}|0\rangle\ +\
                         c_1(t)e^{-i\omega_1 t}|1\rangle\ \ ,
\end{equation}
the equation of motion for the system (atom $+$ laser fields) 
in the rotating frame reads
\begin{eqnarray}
      i\dot{c_e}\ & = & \ (1/2) (2\Delta\  c_e\ +\ \Omega_0\ c_0\ + \Omega_1\ c_1)
\nonumber
\\
      i\dot{c_0}\ & = & \ (1/2)\  \Omega_0\  c_e
\label{eom}
\\
      i\dot{c_1}\ & = & \ (1/2)\  \Omega_1\  c_e
\nonumber
\end{eqnarray}
where we have 
neglected finite lifetimes of the atomic levels.
The corresponding Hamiltonian written in the basis $\{|e\rangle,
|0\rangle,|1\rangle\}$ is
\begin{equation}\label{Hamiltonian}
        H\ =\ \frac{1}{2}\left[
              \begin{array}{ccc}
              2 \Delta & \Omega_0 & \Omega_1\\
              \Omega_0 &   0    & 0\\
              \Omega_1 &   0    & 0
              \end{array}
              \right]\  \  .
\label{rotframHam}
\end{equation}
The eigenenergies of this Hamiltonian are easily found
to be 
\begin{equation}
  E^0 = 0\ \  \ \ \mathrm{and}
                 \ \ \ E^{\pm} =  \frac{1}{2}\left(\Delta\pm
                                \sqrt{\Delta^2 + \Omega_0^2 + \Omega_1^2}
                                                \right)\ \ .
\end{equation}
Note that one of the two zero diagonal elements of the Hamiltonian 
(\ref{rotframHam}) survives {\em independently} of the 
couplings $\Omega_j$ and the offset $\Delta$. The corresponding eigenstate
(which is {\em not} the ground state) reads
\begin{equation}
     |\phi^0\rangle\ = \ \frac{1}{\sqrt{\Omega_0^2+\Omega_1^2}}\
                         (\ \Omega_1\ |0\rangle\ -\ \Omega_0\ |1\rangle\ )\ \ .
\label{darkstate}
\end{equation}
The remarkable property of this state is that it does not contain an 
admixture of the basis vector $|e\rangle$. Therefore it cannot decay
by spontaneous emission from $|e\rangle$ and is  usually called a
`dark state'. The direction of $|\phi^0\rangle$
in the subspace $\{|0\rangle,|1\rangle\}$ 
is given by the (possibly complex)
coupling parameters $\Omega_j$.

Now consider the case of  weakly time-dependent coupling parameters.
If the inverse characteristic scale of this time dependence is smaller than 
the level spacing $\Omega_{\mathrm{eff}}$ of the Hamiltonian (\ref{rotframHam})
\begin{equation}
                 \left|\frac{\dot{\Omega_j}}{\Omega_j}\right|\ll 
                          \Omega_{\mathrm{eff}}\sim|\Omega_j|
\label{adiabatic}
\end{equation}
the adiabatic theorem guarantees that a given state follows the 
time evolution of the coupling coefficients. 
Consequently, the dark state can be rotated 
in the subspace 
$\{|0\rangle,|1\rangle\}$ by slowly changing the values of the couplings.
Again, this is remarkable as the {\em quantum} state 
can be controlled by varying {\em classical} system parameters.

Population transfer from the state $|0\rangle$ to the state $|1\rangle$
can be achieved by applying the so-called counterintuitive scheme.
The system is prepared in the state $|0\rangle$ at vanishing couplings.
Switching on the laser 1 to a finite $\Omega_1$ does not affect the state 
of the system. Now, also 
the laser 0 is slowly switched on (finite $\Omega_0$) while 
$\Omega_1$ is turned off. As can be immediately read off Eq.\ (\ref{darkstate})
the dark state $|\phi^0\rangle$ starts to rotate towards the state $|1\rangle$.
At the end of the switching procedure we have $\Omega_1=0$, finite 
$\Omega_0$, and the final state equals $|1\rangle$.

\section{Analogy for a Cooper-pair box with three islands}

Our aim is to show that Hamiltonians similar to Eq.\ (\ref{rotframHam})       
can be `tailored' in solid-state devices. As an example we consider
a superconducting Cooper-pair box with three islands. The (single-island)
Cooper-pair box is a well-known system as it is a promising candidate
for the practical realization of a solid-state qubit (see, e.g., 
\cite{YuriRMP} and references therein).

\begin{figure}
\begin{center}
\includegraphics[width=.6\textwidth]{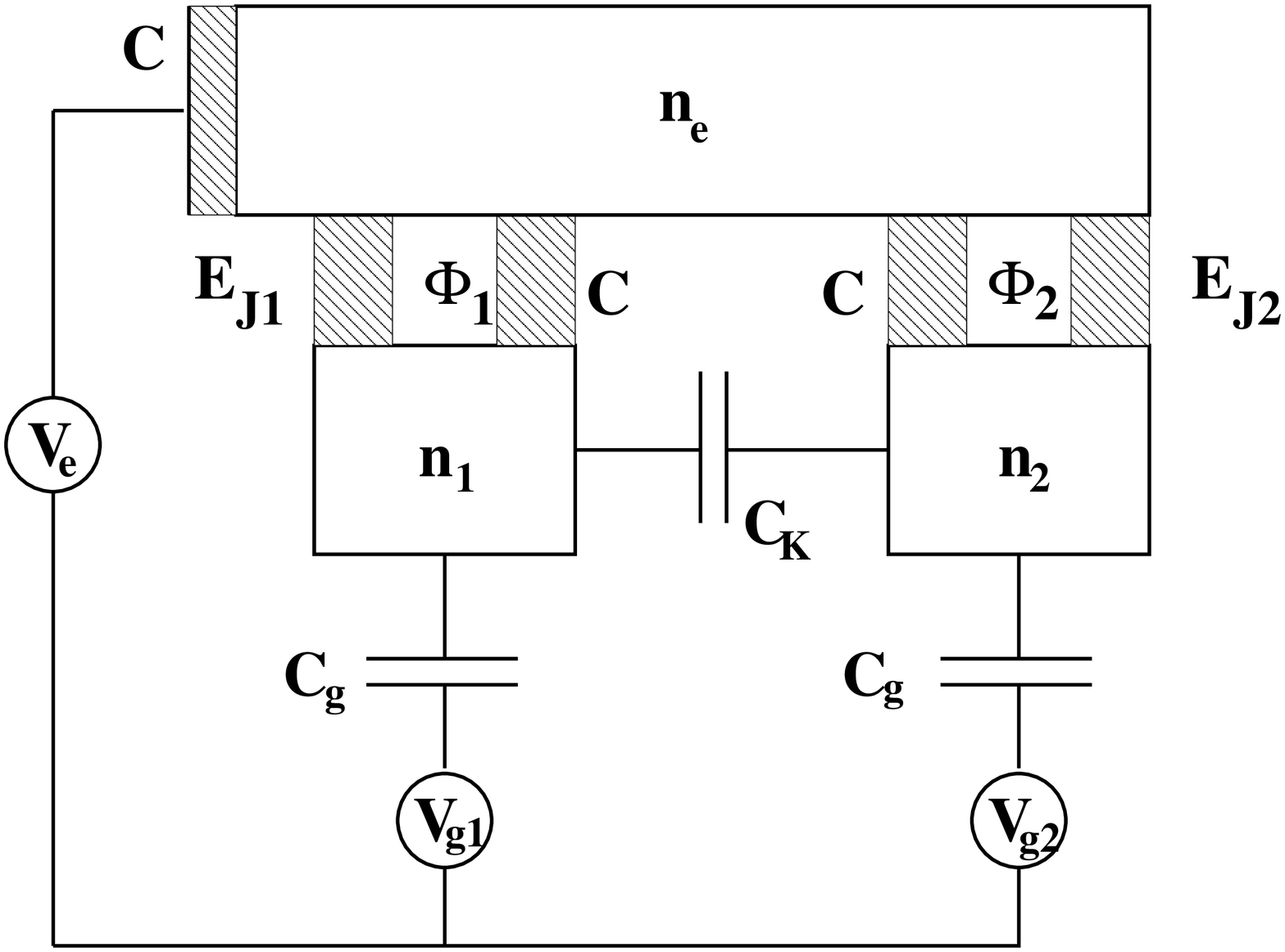}
\end{center}
\caption[]{Circuit for a Cooper-pair box with three islands.
           For simplicity we choose equal gate capacitances $C_g$
           and also equal junction capacitances $C\gg C_g$. The system
           is operated in the charge regime $E_C\gg E_{Jj}$.
           The $\Lambda$ scheme can be recognized by looking at
           the position of the excess Cooper pair (for a detailed explanation,
           see text): $|0\rangle=$ pair in island (1), $|1\rangle=$ pair 
           in island (2), $|e\rangle=$ pair in island (e).
           }
\label{fig2}
\end{figure}

The circuit of a three-island Cooper-pair box is shown in Fig.\ \ref{fig2}.
The superconducting islands (1) and (2) are coupled via tunable Josephson
junctions to the third island (e).  Moreover, there is a capacitive
coupling $C_K$ between the islands (1) and (2) which is required to
generate an appropriate energy level structure of the device.
The junction between island (e)
and superconducting lead serves only to change 
the total charge on the three-island setup.
The SQUID-loop layout of the junctions (1) and (2) makes it possible
to control the coupling energies $E_{J1}$, $E_{J2}$ by means of the
external fluxes $\Phi_1$, $\Phi_2$. The electrostatic potentials of
the islands can be changed through the gate voltage sources $V_{gj}$ ($j=1,2,e$)
which induce offset charges $n_{xj}=C_{gj}V_{gj}/2e$, 
$n_{xe}=CV_e/2e$ on the islands (in units of Cooper-pair charges).

Let us calculate the energy of the circuit in Fig.\ \ref{fig2}
according to classical
electrostatics.
We assume $C_{g1}=C_{g2}=C_g \ll C_1=C_2=C$ and $C_g\ll C_K$.
Then the electrostatic energy can be written in terms of the
number of island charges $n_j$ (more precisely: excess charges) as 
\begin{eqnarray}
   {\cal E}(N,n_1,n_2) & \simeq  &
       E_C\left\{(N-N_x)^2 + 
      \frac{C+C_K}{C+2C_K}\left[(n_1-n_{x1})^2+ (n_2-n_{x2})^2\right]
      \right.
 \nonumber\\
     && \ +\ \left.
      \frac{2C_K}{C+2C_K}(n_1-n_{x1})(n_2-n_{x2})\
                                + \mathrm{const.}\right\}
\label{elenerg}
\end{eqnarray}
where $E_C=(2e)^2/(2C)$ denotes the charging-energy scale
and $N\equiv n_e+n_1+n_2$ is the total
charge number on all three islands (and correspondingly 
\mbox{$N_x\equiv n_{xe}+n_{x1}+n_{x2}$}).
Observe that the number of island charges $n_j$ are {\em discrete} variables.

Assume we choose $N_x\simeq 1$. Then, the charge states with the lowest
electrostatic energy correspond to a total charge of $N=1$.
If we further put $n_{x1}=n_{x2}=0.4$ 
the charge states with $n_{1}=1$, $n_{2}=0$
(and, correspondingly, $n_{1}=0$, $n_{2}=1$) have the same energy.

In summary, we have obtained that, given the offset charges 
mentioned above,  the three charge charge states with the lowest
electrostatic energy are $|0\rangle$ with ``one charge on island 1, no
charge on island 2'', $|1\rangle$ with ``no charge on island 1,
one charge on island 2'', and $|e\rangle$ with zero excess charge
on both islands 1 and 2. 
 All other charge states have energies that 
 are on the order of $E_C$ higher. 
 Note that ${\cal E}(1,1,1)- {\cal E}(1,1,0) > 
                {\cal E}(1,0,0)- {\cal E}(1,1,0) $. 

Now we include also Josephson tunneling in our discussion.
The device is to be operated in the charge regime, that is
\begin{equation}
   \ \ \ \ \ \ \ \       E_{Jj}\ \ll \ E_C \ \ .
\end{equation}
If we are interested in the low-energy dynamics it is sufficient
to consider only the lowest-lying charge states and the relevant
Josephson couplings. This is in complete analogy
with the reasoning for the one-island Cooper-pair box in Ref.\ \cite{YuriRMP}.
The resulting Hamiltonian that describes the quantum dynamics of the circuit 
in Fig.\ \ref{fig2} is
\begin{eqnarray}
 H\ & = &\ {\cal E}(1,0,0)|e\rangle\langle e|
                          +{\cal E}(1,1,0)|0\rangle\langle 0|
                                       +{\cal E}(1,0,1)|1\rangle\langle 1|-
\nonumber
\\
    &&           -(E_{J1}/2)(|e\rangle\langle 0|+|0\rangle\langle e|)
                 -(E_{J2}/2)(|e\rangle\langle 1|+|1\rangle\langle e|)\ \ ,
\label{chargeHam}
\end{eqnarray}
where the first line describes the charging part discussed above, and
the second line is the tunneling Hamiltonian in the three-dimensional
subspace. As it is possible to choose ${\cal E}(1,1,0)={\cal E}(1,0,1)$,
we see that the Hamiltonian \mbox{Eq.\ (\ref{chargeHam})} has exactly the same
structure as the Hamiltonian Eq.\ (\ref{rotframHam}) of the three-level
atom in the rotating frame.

From the mathematical equivalence of the Hamilton operators in
Eqs.\ (\ref{rotframHam}) and (\ref{chargeHam}) we conclude 
that an adiabatic population transfer as described in Section 2
is possible also in the three-island Cooper-pair box. 
Here, switching the coupling parameters means to vary the
Josephson tunnel couplings by tuning the local fluxes $\Phi_1$, $\Phi_2$.
Physically,
the population transfer corresponds to moving the excess Cooper pair
from island 1 to island 2.

\section{Solid-State Applications of Adiabatic Passage}

In this section, we will illustrate further applications of
adiabatic population transfer in solid-state devices.
It is evident from these examples that the method provides 
unexpected solutions to interesting problems, therefore one
might hope that it finds a wider range of condensed-matter applications in
the future. 

\subsection{Non-Abelian Holonomies by Sequences of Adiabatic Population
            Transfers}
It is certainly an interesting problem to demonstrate the existence of
geometric phases and to measure them quantitatively \cite{Shapere}.
Numerous manifestations of geometric phases in physics are 
so-called Berry phases \cite{Berry}. This phase occurs when a {\em non-degenerate}
quantum state carries out a cyclic evolution due to cyclic adiabatic
parameter changes of the Hamiltonian.

The generalization of the Berry phase to {\em degenerate} states 
is the non-Abelian holonomy \cite{WilczekZee84}. 
Consider a quantum system which depends
on an $n$-tuple of parameters $\{\lambda_1,\ldots, \lambda_n\}$
(external fields, etc.), the {
\em control manifold}. Moreover, let the system have a degenerate
subspace which remains degenerate for any parameter point of the
control manifold (this is a rather non-trivial assumption). We
prepare the system in a state that is an element of the degenerate
subspace and perform a cyclic adiabatic evolution of the Hamiltonian
along a closed contour in the control manifold.

While a non-degenerate state returns to the initial state (times a phase
factor) at the end of the cyclic evolution (a consequence of the adiabatic 
theorem), a state in the degenerate subspace will, in general,  experience
a rotation $U$ within the degenerate subspace. This rotation is called a
non-Abelian holonomy. Mathematically $U$ is given by a path-ordered
integral along the contour $C$
\begin{equation}
    U\ =\ {\cal P}\ \exp{\oint_C \chi\ \mathrm{d}\lambda}
\end{equation}
where $\chi$ is the (matrix-valued) Wilczek-Zee connection~\cite{WilczekZee84}. 
In general this expression
is rather difficult to evaluate. Therefore, it is an interesting
question whether one
can find contours in the control manifold for which one can immediately 
see the corresponding holonomy $U$. This is of relevance also for an
experimental demonstration of non-Abelian holonomies.

\begin{figure}
\begin{center}
\includegraphics[width=.8\textwidth]{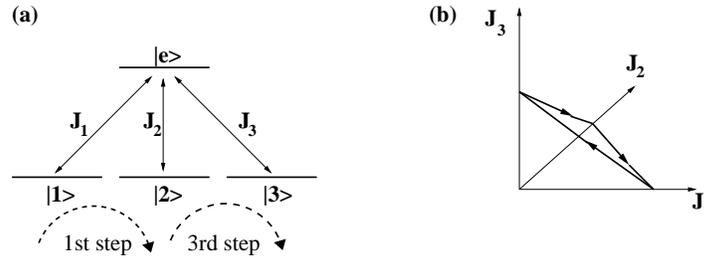}
\end{center}
\label{fig3}
\caption[]{{\bf (a)} Four-level scheme for the generation of non-Abelian
               holonomies. The energies of the levels (i.e., the
               diagonal elements of the 4$\times$4 Hamiltonian
               analogous to Eq.\ (\ref{rotframHam})) are $E_e=\varepsilon$
               and $E_1=E_2=E_3=0$. {\bf (b)} Contour in the control manifold
               described in the text. The starting point is 
               $J_1=J_3=0$, $J_2$ finite.
           }
\end{figure}

Adiabatic passage provides an answer to this question that has been
found in the context of holonomic quantum computation 
\cite{Unanyan99,Duan01,NonAbel}.
Consider a four-level scheme with states $|e\rangle$, $|1\rangle$, 
$|2\rangle$, $|3\rangle$ (see Fig.\ \ref{fig3}a) where the control manifold is
given by the coupling parameters $\{J_1,J_2,J_3\}$. This system can 
be realized, e.g., in a superconducting nanocircuit with three islands
in analogy with the circuit in Fig.\ \ref{fig2} (see also Ref.\ \cite{NonAbel}
where a similar system has been studied).

It is easy to show
that this system has a two-dimensional degenerate subspace for any
parameter configuration in the control manifold. The system is
prepared in the state $|1\rangle$ and $J_1=0$, $J_2=
\cal J$, $J_3=0$. Now we perform a 3-step sequence
of adiabatic passages (cf.\ also Fig.\ \ref{fig3}b):
\begin{itemize}
     \item[] {\em 1st step}: switch off $J_2$ while $J_1$ is switched on \\
          \hspace*{35mm}$\longrightarrow$ state of system changes 
                                          $|1\rangle\rightarrow|2\rangle$
     \item[] {\em 2nd step}: switch off $J_1$ while $J_3$ is switched on\\     
          \hspace*{35mm}$\longrightarrow$ state of system does not change
     \item[] {\em 3rd step}: switch off $J_3$ while $J_2$ is switched on\\     
          \hspace*{35mm}$\longrightarrow$ state of system changes 
                                          $|2\rangle\rightarrow|3\rangle$
\end{itemize}

We see that, while a closed contour is described
in the control manifold, the state of the system is rotated from 
state $|1\rangle$ to state $|3\rangle$ (possibly times a phase factor). 
That is, we have found a
simple non-Abelian holonomy which can be generated experimentally
in a straightforward manner.

\subsection{Coupled Quantum Dots}
\begin{figure}
\begin{center}
\includegraphics[width=.6\textwidth]{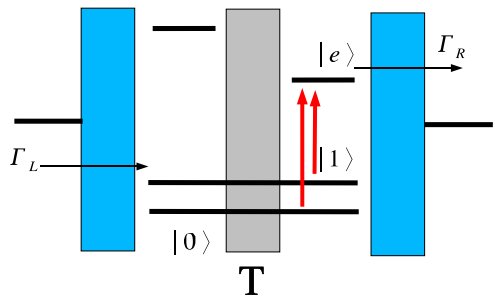}
\end{center}
\caption[]{Dark states and STIRAP through two tunnel ($T$) coupled dots in the Coulomb blockade
regime with hybridized states $|0\rangle$, $|1\rangle$. Electrons tunnel in from the left,
are photo-pumped to the  excited state $|e\rangle$, and tunnel out to the right.
           }
\label{fig4}
\end{figure}

Another solid-state application is the realization of dark states and adiabatic passage
in coupled quantum dots. The original proposal \cite{Tobias1} with two coupled
quantum dots in the strong Coulomb blockade regime is actually quite close to three-level
systems in atoms, with the additional possibility to test the effect (and its modifications) in
electronic transport, cf. Fig. \ref{fig4}. The dark state \ Eq.~(\ref{darkstate}) appears in the
form of a sharp anti-resonance in the {\em stationary current} through a double dot as a function of 
the `Raman detuning', i.e. the detuning difference of the two classical laser (or microwave) fields. 
The half-width of the anti-resonance can then be used to extract valuable information, such
as the relaxation and dephasing times of tunnel coupled dot-ground state superpositions, 
from transport experiments.

Using two time-dependent Stokes and pump pulses, an extension of the counterintuitive STIRAP scheme
has subsequently been suggested for this configuration \cite{Tobias1}, taking into account
a finite decoherence rate $\gamma$ due to, e.g. electron-phonon coupling in the dots. 
In principle, $\gamma$ can be obtained from monitoring the time-dependent 
electronic current through the dots, which however is quite weak for small $\gamma$, when the dot
is essentially trapped in the dark superposition of $|0\rangle$ and $|1\rangle$. 
An alternative way is to apply a {\em second} pair of simultaneous pulses with 
amplitude ratios that give either zero or full current in the coherent case $\gamma=0$. The deviation
from the `zero/full' current situations then gives rise for a current `contrast' from which
$\gamma$ can be extracted.

Another adiabatic scheme that completely avoids the use of lasers or microwaves 
has been introduced in quantum dots which are coupled by slowly varying static tunnel barriers. 
In fact, the rotating wave approximation in the original (optical) population transfer scheme
leads to time-independent (or slowly parametric) Hamiltonians
like \ Eq.~(\ref{Hamiltonian}), where fast terms are already transformed away. This makes it obvious that one 
can start from time-dependent tunnel couplings right from the beginning. 
In the simplest realization \cite{Tobias2}, one considers {\em three} single level dots
$L$, $C$, $R$ in a line, with two time-dependent couplings $T_i(t)$ between $C$ and $i=L,R$ which are then 
switched on and off with a time-delay as in the STIRAP scheme. 
The resulting adiabatic transfer of charge from the left
to the center to the right can then essentially be understood in terms of level-crossings of the
(instantaneous) three eigenvalues  of the energy. If the tunnel coupling remains small but finite in the 
`off' periods, these level crossings become anti-crossings. Already for 
the two-level system in a double quantum dot, one realizes that one actually has to deal with 
(dissipative) Landau-Zener
tunneling between curves on energy surfaces in the parameter space of the problem \cite{Tobias3}.
This type of `pumping', which occurs in Hilbert spaces that are essentially cut down to very small
dimensions due to strong correlations, is the opposite limit of the `usual' adiabatic pumping 
in large, non-interacting mesoscopic systems \cite{Brouwer98}.
 
\section{Conclusions}

We have outlined the basic ideas of adiabatic passage and
several examples of its application in solid-state devices
that can be realized in superconductor as well as in
semiconductor systems.
The examples showed that adiabatic passage-like techniques
may be used to generate simple charge transfers, to
detect non-Abelian holonomies and to control charge transport 
properties (in particular also to design an alternative 
kind of charge pumps). In our opinion, it is obvious even 
from these simple examples that the method of adiabatic
passage has significant potential for applications
in solid-state physics and deserves appropriate attention
also in this field. 

The authors gratefully acknowledge stimulating discussions
with L.\ Faoro, R.\ Fazio, A.\ Kuhn, F.\ Renzoni, and T.\ Vorrath.

%


\begin{thebibliography}{8.}
\addcontentsline{toc}{section}{References}

\bibitem{BergBuch} K.\ Bergmann and B.W.\ Shore: Coherent
        population transfer. In H.L.\ Dai and R.W.\ Field (Eds.)
        {\itshape Molecular dynamics and stimulated emission 
                  pumping}, p.\ 315.
 (World Scientific, Singapore, 1995)

\bibitem{BergRMP} K.\ Bergmann, H.\ Theuer, and B.W.\ Shore:
                  Coherent Population Transfer Among Quantum
                  States of Atoms and Molecules,
Rev.\ Mod.\ Phys.\ \textbf{70}, 1003 (1998).

\bibitem{Tobias1} T. Brandes and F. Renzoni: 
       Current switch by coherent
       trapping of electrons in quantum dots,
       Phys. Rev. Lett. {\bf 85}, 4148 (2000); T. Brandes, F. Renzoni,
       and R. H. Blick: 
       Adiabatic steering and determination
       of dephasing rates in double-dot qubits,
       Phys. Rev. B {\bf 64}, 035319 (2001). 

\bibitem{Tobias2} F. Renzoni and T. Brandes: 
       Charge transport through quantum
       dots via time-varying tunnel couplings,
       Phys. Rev. B {\bf 64}, 245301 (2001).


\bibitem{NonAbel} L.\ Faoro, J.\ Siewert, and R.\ Fazio:
        Non-Abelian Holonomies, Charge Pumping and Quantum Computation
        with Josephson Junctions,
        Phys.\ Rev.\ Lett.\ {\bf 90}, 028301 (2003).       

\bibitem{Alec} M.H.S.\ Amin, A.Yu.\ Smirnov, and A.\ Maassen v.d.\ Brink:
        Josephson-phase qubit without tunneling,
        Phys.\ Rev.\ B {\bf 67}, 100508(R) (2003).


\bibitem{Scully} M.O.\ Scully and M.S.\ Zubairy: {\itshape Quantum Optics} 
(Cambridge Univ.\ Press, Cambridge 1997)

\bibitem{YuriRMP} Yu.\ Makhlin, G.\ Sch\"on, and A.\ Shnirman:
   Quantum-state engineering with Josephson-junction devices,
   Rev.\ Mod.\ Phys.\ {\bf 73}, 357 (2001).

\bibitem{Shapere} 
        A.\ Shapere and F.\ Wilczek  (Eds.): 
        {\itshape Geometric phases in physics},
        (World Scientific, Singapore, 1989).


\bibitem{Berry} 
        M.V.\ Berry, 
        Quantum Phase Factors Accompanying
        Adiabatic Changes, 
        Proc.\ Roy.\ Soc.\ A {\bf 392}, 45 (1984).

\bibitem{WilczekZee84} 
        F.\ Wilczek and A.\ Zee, 
        Appearance of Gauge Structure in Simple Dynamical Systems,
        Phys.\ Rev.\ Lett.\ {\bf 52}, 2111 (1984).

\bibitem{Unanyan99} R.G.\ Unanyan, B.W.\ Shore, and K.\ Bergmann,
        Laser-driven population transfer in four-level atoms: 
        Consequences of non-Abelian geometrical adiabatic phase factors,
        Phys.\ Rev.\ A {\bf 59}, 2910 (1999).

\bibitem{Duan01} L.-M.\ Duan, J.I.\ Cirac, and P.\ Zoller,
        Geometric Manipulation of Trapped Ions for Quantum Computation,
        Science {\bf 292}, 1695 (2001).



\bibitem{Tobias3} T.\ Brandes and T.\ Vorrath, 
        Adiabatic transfer of electrons in coupled quantum dots,
        Phys.\ Rev.\ B {\bf 66}, 075341 (2002).

\bibitem{Brouwer98} P.W.\ Brouwer: Scattering approach to parametric pumping, 
         Phys.\ Rev.\ B {\bf 58}, 10135 (1998).


\end{thebibliography}
\end{document}